\documentclass[12pt]{article}
\pdfoutput=1
 \usepackage[nottoc]{tocbibind}
\usepackage[colorlinks=true,linkcolor=blue,urlcolor=magenta, citecolor=blue]{hyperref}
\usepackage{fullpage}
\usepackage{amssymb,graphicx}
\usepackage{amsmath}
\usepackage[margin=0.5cm]{caption}
\usepackage{hyperref} 
\usepackage{cite}
\usepackage{upgreek}

\def\d{\partial}
\def\l{\left(}
\def\r{\right)}

\newcommand{\be}{\begin{equation}}
\newcommand{\ee}{\end{equation}}
\newcommand{\bea}{\begin{eqnarray}}
\newcommand{\eea}{\end{eqnarray}}
\newcommand{\bg}{\begin{gather}}
\newcommand{\eg}{\end{gather}}
\newcommand{\bseq}{\begin{subequations}}
\newcommand{\eseq}{\end{subequations}}

\begin{document}
\baselineskip=15.5pt
\begin{titlepage}
\begin{center}
{\Large\bf   $T\bar{T}$ Partition Function from Topological Gravity}\\
\vspace{0.5cm}
{ \large
Sergei Dubovsky$^{a}$, Victor Gorbenko$^b$, and Guzm\'an Hern\'andez-Chifflet$^{a,c}$ 
}\\
\vspace{.45cm}
{\small  \textit{  $^a$Center for Cosmology and Particle Physics,\\ Department of Physics,
      New York University\\
      New York, NY, 10003, USA}}\\ 
      \vspace{.1cm}
      \vspace{.25cm}      
      {\small \textit {{$^b$Stanford Institute for Theoretical Physics,\\ Stanford University, Stanford, CA, USA}}}
      \\
       \vspace{.1cm}
      \vspace{.25cm}    
      {\small  \textit{   $^c$Instituto de F\'isica, Facultad de Ingenier\'ia,\\ Universidad de la Rep\'ublica,\\
      Montevideo, 11300, Uruguay}}\\ 
\end{center}
\begin{center}
\begin{abstract}
The $T{\bar T}$ deformation of a relativistic two-dimensional theory results in a solvable gravitational theory. Deformed scattering amplitudes can be obtained 
from coupling the undeformed theory to the flat space Jackiw--Teitelboim (JT) gravity. We show that the JT description is applicable and useful also in  finite volume. Namely, we calculate the torus partition function of an arbitrary matter theory coupled to the JT gravity, formulated  in the first order (vielbein) formalism. The first order description provides a natural set of dynamical clocks and rods for this theory, analogous to the target space coordinates in string theory. These dynamical coordinates play the role of relational 
observables allowing to define a torus path integral for the JT gravity. The resulting partition function is one-loop exact and reproduces the $T\bar{T}$ deformed finite volume spectrum.
\end{abstract}
\end{center}
\end{titlepage}
\newpage
\section{Introduction} 
Consider a relativistic quantum field theory in two spacetime dimensions. It can be fully characterized by its  $S$-matrix elements $S(p_i)$, where $p_i$ is a set of on-shell momenta, which are all taken to be incoming. It was observed in \cite{Dubovsky:2013ira} that for any $S$ one can construct a one-parametric family of new consistent dressed $S$-matrices of the form
\be
\label{USU}
\hat{S}=USU\;.
\ee
Here $U$ is a unitary operator acting in the Fock space as
\be
\label{U}
U|\{p_i\}\rangle=e^{i\ell^2\sum_{i<j}p_i*p_j}|\{p_i\}\rangle\;,
\ee
where
\[
p_i*p_j=\epsilon_{\alpha\beta}p_i^\alpha p_j^\beta
\]
and the momenta in (\ref{U}) are ordered according to their rapidities. $\ell^2$ is the deformation parameter. The first and the simplest example of this construction is the worldsheet $S$-matrix for critical strings, which can be obtained by deforming a theory of 24 free massless bosons (with an obvious generalization to the superstring case) \cite{Dubovsky:2012wk}. A dressed theory exhibits gravitational features and gives rise to a novel asymptotic UV behavior,  dubbed asymptotic fragility. 

As explained in \cite{Caselle:2013dra,Smirnov:2016lqw,Cavaglia:2016oda},
the existence of this deformation is related to certain remarkable properties of the ``$T\bar{T}$" operator \cite{Zamolodchikov:2004ce}
\be
\label{TTbar}
T\bar{T}\equiv {1\over 2}\l T_{\alpha\beta}T^{\alpha\beta}-T^{\alpha 2}_\alpha\r\;.
\ee
Namely, a family of $S$-matrices (\ref{USU}) can be interpreted as a trajectory in field theory space such that its tangent vector at each point is set by the $T{\bar  T}$ operator. This description allows to calculate the effect of the deformation (\ref{USU}) on the finite volume spectrum of a theory compactified on a circle of  circumference $R$.
Namely, the energies of individual Kaluza--Klein modes  satisfy the following differential equation,
 \be
 \label{burgers}
 \d_{\ell^2} E_n={1\over 2}\l{P_n^2\over R}+E_n\d_R E_n\r\;,
 \ee
where $E_n$ is the energy of a state labeled by $n$ and $P_n$ is its total momentum, which is equal to
\[
P_n={2\pi k_n\over R}\;,
\]
for some integer $k_n$.
A further important observation was made in \cite{Cardy:2018sdv}, where it was pointed out that the non-linear ``hydrodynamical" equation (\ref{burgers}) for the energy levels implies that the torus partition function $Z_{T\bar T}$ of a deformed theory satisfies a 
{\it linear} diffusion-like equation. Namely, one finds
\be
\label{cardy}
\d_{\ell^2}\Psi={1\over 2}(\d_{L_2}\d_{L'_1}-\d_{L_1}\d_{L'_2}) \Psi\;,
\ee
where the torus is described as a parallelogram with vertices at 
\be
\label{LLptorus}
A=(0,0),\;B=(L_1,L_2),\;C=(L'_1,L'_2),\;D=(L_1+L'_1,L_2+L'_2)\;,
\ee
\begin{figure}[t!]
  \begin{center}
        \includegraphics[height=8cm]{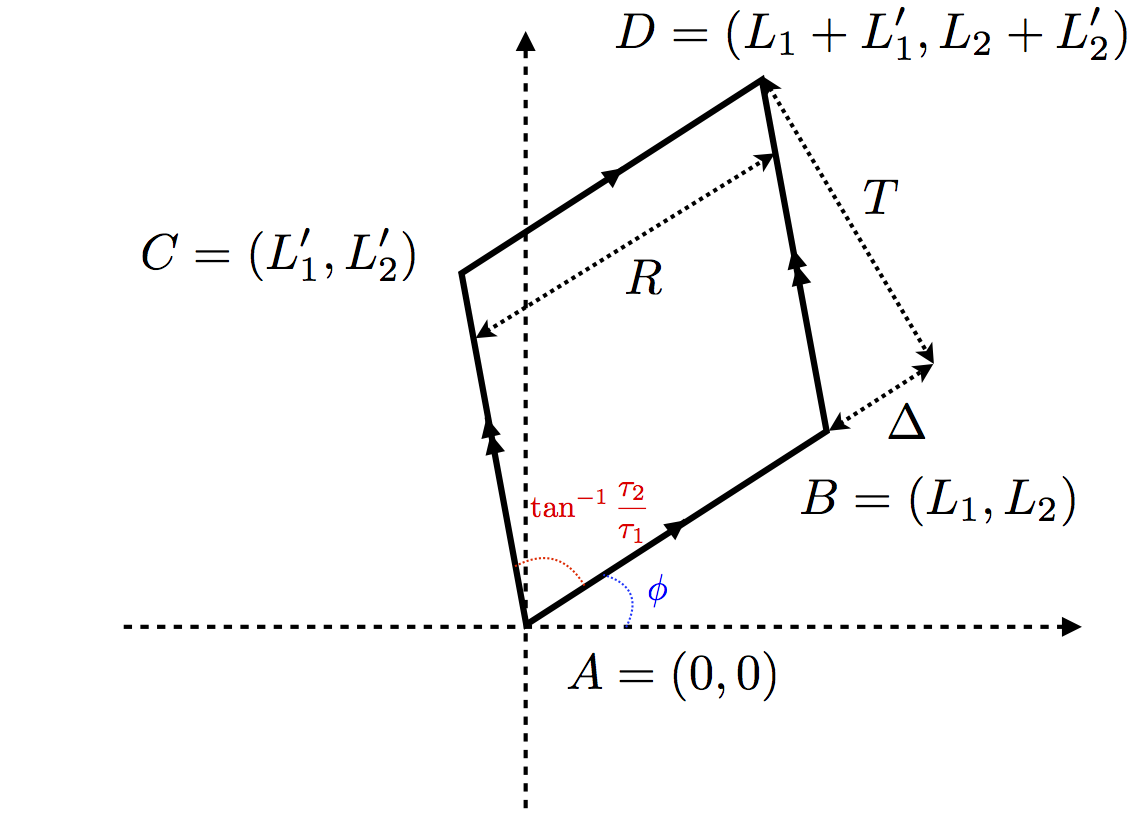} 
           \caption{ A torus can be represented as a parallelogram with identified opposite sides. Its angle is determined by modular
           parameters, $\measuredangle BAC=\arctan{\tau_2\over \tau_1}$. The overall orientation of the parallelogram is determined by the angle $\phi$, which does not affect the internal geometry of the torus.}
        \label{fig:ABCD}
    \end{center}
\end{figure}
as shown in Fig.~\ref{fig:ABCD}. Here $\Psi$ is related to the partition function via
\be
\label{psi}
\Psi={\cal A}^{-1}Z_{T\bar T}\;.
\ee
where
\be
\label{calA}
{\cal A}=L_1L'_2-L'_1L_2
\ee
is the area of the torus\footnote{This factor of ${\cal A}$ was missed in an earlier version of \cite{Cardy:2018sdv}. In what follows it comes out from the proper zero mode measure in the gravitational path integral. See Appendix~\ref{app:diffusion} for the direct derivation of (\ref{cardy}) from (\ref{burgers}).}. 

We see that the $T{\bar T}$ description of the deformation (\ref{USU}) is definitely useful and provides  new important
insights into its properties. However, it is rather different from the conventional ways to construct quantum theories,
especially given that $T{\bar T}$ is an irrelevant operator.
The nature of the $T{\bar T}$ deformed theories would be much more transparent if they could be defined in a conventional path integral formalism. In particular, certain properties, such as modular invariance, would be made more explicit. 

A concrete proposal for such a description was made in \cite{Dubovsky:2017cnj}. Namely, it was proven there that the dressed $S$-matrix 
can be obtained by coupling an undeformed field theory to a gravitational sector described by the flat space Jackiw--Teitelboim (JT) gravity. The full action for the deformed theory then takes the following form,
\be
\label{action}
S_{T\bar{T}}=S_0(g_{\alpha\beta},\psi)+\int \sqrt{-g} (\varphi R-\Lambda)\;.
\ee
Here $S_0(g_{\alpha\beta},\psi)$ is the action of the original field theory with a minimal coupling to the dynamical metric $g_{\alpha\beta}$ and no direct coupling to the field $\varphi$. 
The deformation parameter $\ell^2$ is determined by the vacuum energy $\Lambda$,
\be
\label{ell}
\ell^2=-{2\over \Lambda}\;.
\ee

The gravitational sector in (\ref{action}) is purely topological. It does not bring in any new local propagating degrees of freedom, and 
the JT dilaton $\varphi$ plays a role of a Lagrange multiplier forcing the metric to stay flat. 
Large diffs are the only new dynamical degrees of freedom introduced in (\ref{action}).
Conventionally, topological gravity is  expected to couple to topological sigma models  and
the resulting physical observables are certain topological invariants \cite{Witten:1989ig,Witten:1990hr,Dijkgraaf:2018vnm}.
  Somewhat surprisingly, we find here that topological gravity can also be consistently coupled to conventional field theories and this coupling induces a non-trivial modification of dynamical observables (scattering amplitudes and  finite volume energies). 
  
It is worth noting, however, that \cite{Dubovsky:2017cnj} fell short of deriving the finite volume spectrum (\ref{burgers}), apart from the critical case when the matter is described by a $c=24$ CFT. Related to this, \cite{Dubovsky:2017cnj} made a very limited use of the JT path integral.
In the critical case one can use the Polyakov description instead of the JT one, and the calculation reduces
to the vacuum string amplitude in the sector with two windings. 

The main purpose of the present paper is to fill in this gap and to derive the finite volume spectrum (\ref{burgers}) from the JT path integral for a general matter theory. This calculation
  solidifies the equivalence of the $T{\bar T}$ deformation to the flat space JT gravity. Perhaps an even more interesting aspect of this calculation is that it offers a possibility to explore to what extent the path integral over metrics provides an adequate definition of a gravitational theory in a concrete fully calculable setup. Of course, as a price for calculability, the  gravitational sector here is extremely simple and does not describe any local degrees of freedom.

The rest of the paper is organized in the following way. In section~\ref{sec:Smatrix} we describe the first order (vielbein) formulation of the JT gravity (\ref{action}). In infinite volume this formulation is completely equivalent to (\ref{action}). However, as we explain, it makes certain important symmetries of (\ref{action}) more manifest. As a result it provides a  natural starting point to put the theory in a finite volume and allows to resolve the previously 
encountered difficulties. The torus partition function of the resulting theory is calculated in section~\ref{sec:finite}, with some gory details postponed till Appendices~\ref{app:gory1} and \ref{app:gory2}. The result is one-loop exact and reproduces the $T\bar{T}$ deformed partition function, as expected from the equivalence of the JT $S$-matrix to (\ref{USU}).
We present our conclusions in section~\ref{sec:last}.

\section{First Order Gravitational Action and the $S$-matrix}
\label{sec:Smatrix}
The key step in the derivation of the JT $S$-matrix presented in \cite{Dubovsky:2017cnj} is to identify dynamical coordinates $X^a$, $a=0,1$, which play the role of relational observables. These variables are analogs of target space coordinates in a critical string theory. In fact, the argument of  \cite{Dubovsky:2017cnj} is a direct generalization of the derivation of the string worldsheet $S$-matrix in the Polyakov formalism as presented in \cite{Dubovsky:2016cog}\footnote{To be precise, we refer here to the first out of three derivations presented in \cite{Dubovsky:2017cnj}, which can be found in Section 2 of that article. The remaining two derivations also rely on a careful identification of dynamical physical coordinates, but perhaps less explicitly.}.

In the critical string case the calculation of the worldsheet partition function is straightforward. It reduces to a calculation of the one-loop vacuum amplitude in the presence of windings for both $X^0$ and $X^1$. However, extending this prescription to a general JT case turns out somewhat challenging. The problem can be traced to the definition of the dynamical coordinates in \cite{Dubovsky:2017cnj}. On an infinite plane these were defined only in  the conformal gauge, where they take form
\be
X^\pm=-\ell^2\d_\mp\varphi\;. \label{eq_Xpm}
\ee
It is not clear that coordinates defined this way are scalars.
As a result, shift symmetries $X^\pm\to X^\pm+const$ are not manifest before the gauge fixing, which makes it hard to introduce windings.
To make things worse, conformal gauge fixing is quite different on a torus, due to the presence of moduli. 
Another face of the problem is that  vacuum configurations of the JT dilaton field on a plane
\[
\varphi=-\ell^{-2} (\sigma^++a)(\sigma^-+b)+c\;,
\]
do not look well suited to be compactified on a torus (here  $\sigma^\pm$ are light cone coordinates, and $a$, $b$ and $c$ are integration constants).

As we will see now all these problems get resolved by switching to the first order (vielbein) description of the JT gravity.
A first order description also naturally arises in the holographic construction of the $T\bar{T}$ RG flow \cite{McGough:2016lol}.
 One may be puzzled  why an apparently equivalent reformulation of a theory should make  life easier. An answer could be that the two formulations are indeed completely equivalent, and the first order formalism just provides some technical advantages. Alternatively, it may be that the two descriptions are only equivalent in infinite volume. We will come back to this question later. For now let us introduce the first order formalism and see where it leads us.

Let us separate the total action (\ref{action}) into a sum of gravity and matter contributions,
\[
S_{T\bar{T}}=S_{JT}+S_m\;,
\]
where
\be
\label{Sgr}
S_{JT}=\int \sqrt{-g} \varphi R
\ee
and
\be\label{Sm}
S_m=S_0(g_{\alpha\beta},\psi)-\Lambda\int \sqrt{-g} \;.
\ee
Note that we assigned the vacuum energy to be a part of the matter action.
In the first order formalism one trades the path integral over metrics for the path integral over the vielbein (``dyad") $e_{a\alpha}$, spin connection $\omega_\alpha$ and a pair of Lagrange multipliers,
$\lambda^a$. Here $\alpha,\beta=0,1$  are spacetime tensor indices, and $a,b=1,2$ label components of the fundamental representation of the internal Lorentz  gauge group $SO(1,1)$.
The physical metric $g_{\alpha\beta}$ is expressed through the vielbein in the standard way,
\[
g_{\alpha\beta}=e_{a\alpha} e_{b\beta}\eta^{ab}\;,
\]
where
\be
\eta_{\alpha\beta}=\l\begin{array}{cc} -1&0\\
0&1
\end{array}\r
\ee
is the Minkowski metric.
In particular, the vacuum energy term turns into
\be
\label{evac}
-\Lambda\int \sqrt{-g}=-{\Lambda\over 2}\int \epsilon^{\alpha\beta}\epsilon^{ab}e_{a\alpha}e_{b\beta}\;.
\ee
The gravitational part of the action is written as
\be
\label{SJT}
S_{JT}=\int \epsilon^{\alpha\beta}\l
\lambda^a
\l\d_\alpha e_{a\beta}-\epsilon_a^{\;\;b}\omega_\alpha e_{b\beta} \r+\varphi \d_\alpha\omega_\beta\r\;.
\ee
In this formalism a variation w.r.t. $\lambda^a$ enforces  the metricity constraint for the spin connection\footnote{Here we refer to $\omega_\alpha$ as the spin connection, although usually this term is used for $\epsilon_a^{\;\;b}\omega_\alpha$.}.
The JT dilaton $\varphi$ is, as before, a Lagrange multiplier which now sets the curvature of the spin connection to zero. Its value can be fixed from the $\omega_\alpha$ field equation,
\be
\label{phieq}
\d_\alpha\varphi=\epsilon_{ab}\lambda^ae^b_\alpha\;.
\ee
Hence we can integrate out $\varphi$ and at the same time replace  $\omega_\alpha$ by a flat connection
\[
 \omega_\alpha=\d_\alpha\omega\;.
 \]
Then the action turns into 
\[
S_{JT}=\int \epsilon^{\alpha\beta}\lambda O^{-1}(\omega)\d_\alpha(O(\omega) e_\beta)
\]
where $O(\omega) =  e^{\omega\epsilon}$ is a boost (or, depending on a signature, a rotation) matrix which transforms as
\[
O(\omega)\to O(\omega) g^{-1}
\]
under internal Lorentz (Euclidean rotation) gauge transformation $g$. 
In other words, $\omega$ plays a role of the Stueckelberg field of the internal Lorentz symmetry. As usual with Stueckelberg fields, it is consistent to fix the unitary gauge by setting 
\[
\omega=0
\]
at the level of the action\footnote{In doing so we are neglecting a possibility of a non-trivial holonomy for the spin connection. It does not seem to be relevant for the purposes of the present paper. It will be interesting to understand whether it ever plays a role for more general questions which one may ask about this setup.}.
Then, after performing an additional integration by parts, we finally arrive at the following form of the JT action,
\be
\label{simplified}
S_{JT}=-\int\epsilon^{\alpha\beta}\d_\alpha \lambda^a e_{a\beta}\;.
\ee
The flat space JT gravity is very simple to start with, so it is amusing that it can be further simplified down to (\ref{simplified})! Note that in this formulation reparametrization invariance is the only remaining gauge symmetry;
only the global part of the internal Lorentz symmetry has survived. 
A clear advantage of this description from the viewpoint of the discussion around equation \eqref{eq_Xpm} is that we managed to make manifest two global translational symmetries 
\be
\label{lsym}
\lambda^a\to \lambda^a+const\;.
\ee
This suggests that the Lagrange multipliers $\lambda^a$ (which are gauge invariant scalars now) should be directly related to the dynamical coordinates.
To see the exact relation let us inspect the vielbein field equations, which take the following form 
\be
\label{eleq}
{\epsilon^{\alpha\beta}\over \sqrt{-g}}\d_\alpha \lambda^a=T^{\alpha\beta}e_\alpha^a\;,
\ee
where
\[
T_{\alpha\beta}=-{2\over\sqrt{-g}}{\delta S_m\over \delta g^{\alpha\beta}}
\]
is the matter energy-momentum tensor, which also includes the vacuum energy. 

From now on we can proceed in a complete analogy with the derivation presented in Section 2 of \cite{Dubovsky:2017cnj}.
Namely, on an infinite plane we can fix the conformal gauge where the non-vanishing components of the vielbein $e^a_\alpha$ are (in the light cone coordinates)
\begin{equation}
e_{\pm}^\pm=e^{\Omega\pm \phi}\;.\label{eq_epmlc}
\end{equation}
Variation of (\ref{simplified}) w.r.t. $\lambda^a$ implies that both $\Omega$ and $\phi$ are constant, which without a loss of generality
can be set to zero on a plane  by an appropriate choice of boundary conditions at infinity.
As a result (\ref{eleq}) turns into
\begin{gather}
\label{Y-}
\d_+X^-=-{T_{++}\over \Lambda}\;,\\
\label{Y+}
\d_-X^+=-{T_{--}\over \Lambda}\;,\\
\label{Yth}
\d_{+}X^+=\d_{-}X^-={T_{+-}\over \Lambda}\;.
\end{gather}
where we introduced 
\be
\label{Xa}
X^a=\Lambda^{-1}\epsilon^a_{\,\;b}\lambda^b\;.
\ee
Dynamical coordinates identified in \cite{Dubovsky:2017cnj} satisfy exactly the same equations, so that (\ref{Xa}) provides a desired gauge invariant relation between the Lagrange multipliers $\lambda^a$'s and dynamical clocks and rods (or, equivalently, target space coordinates). In particular, in the Minkowski vacuum one finds
\[
X^a=\sigma^a\;,
\]
which is invariant under the diagonal combination of spacetime translations and of global shifts (\ref{lsym}).

The rest of the $S$-matrix derivation proceeds verbatim as presented in  \cite{Dubovsky:2017cnj}. Namely, one treats (\ref{Y-}), (\ref{Y+}), and (\ref{Yth}) as Heisenberg equations which determine operators $X^a$ in terms of undressed field theory operators. Using these operators in the mode decomposition of asymptotic states one arrives at the gravitational dressing formula (\ref{USU}) for the $S$-matrix.
Note that as a consequence of  (\ref{phieq}) on an infinite plane dynamical coordinates (\ref{Xa}) agree on-shell with those introduced in \cite{Dubovsky:2017cnj}. 
The advantage of the present formalism is that it provides a gauge invariant definition of coordinates, which also makes the corresponding global translational symmetry manifest. This is all what we need to calculate the partition function on a torus, which will be our next step.

\section{Finite Volume Spectrum} 
\label{sec:finite}
To summarize the results of the previous section, we arrived at the following form for the action describing the $T\bar{T}$ deformation,
\be
\label{TTaction}
S_{T\bar{T}}={\Lambda\over 2}\int\epsilon^{\alpha\beta}\epsilon_{ab}\l\d_\alpha X^a -e^a_\alpha\r\l\d_\beta X^b- e^b_\beta \r+S_0(\psi,g_{\alpha\beta})\;,
\ee
where $S_0$ is the action of an undeformed theory. 
When writing (\ref{TTaction}) we added a total derivative in the form
\be
\label{dXdX}
{\Lambda\over 2}\int\epsilon^{\alpha\beta}\epsilon_{ab}\d_\alpha X^a\d_\beta X^b
\ee
to the action we had before. This total derivative does not affect infinite volume scattering;  we will see that on a torus it sets to zero the infinite volume vacuum energy of a deformed theory, in agreement with what one gets from (\ref{burgers}).

From now on we work in  the Euclidean signature, which explains the sign flip for the vacuum energy in (\ref{TTaction}) as compared to (\ref{Sm}). The  metric is
\[
g_{\alpha\beta}=e_\alpha^ae_\beta^b\delta_{ab}\;.
\]
\subsection{Evaluation of the Partition Function}
We can proceed now with calculating the  partition function $Z_{JT}$ of the JT gravity similarly to how it was done  in \cite{Dubovsky:2017cnj} for the critical case.
Namely, the partition function is given by the path integral of the form
\be
\label{ZTT}
Z_{JT}=\int {
{\cal D}e
{\cal D} X
{\cal D}\psi
\over V_{diff}
}
e^{-S_{T\bar{T}}}=\int {
{\cal D}e{\cal D} X\over V_{diff}
}e^{-{\Lambda\over 2}\int\epsilon^{\alpha\beta}\epsilon_{ab}\l\d_\alpha X^a -e^a_\alpha\r\l\d_\beta X^b- e^b_\beta \r}
Z_0(g_{\alpha\beta})\;.
\ee
Here $V_{diff}$ is the volume of the reparametrization group. The ${\cal D} e$ integral in (\ref{ZTT}) goes over all vielbeins on a ``worldsheet" with a topology of a torus. The ${\cal D}X$ integral is performed over mappings of the worldsheet torus into the ``target space" torus (\ref{LLptorus}). The index of these mappings is restricted to one ({\it i.e.}, there is a unit winding along each of the cycles).
$Z_0(g_{\alpha\beta})$ is the undeformed partition function on a worldsheet torus. For a general  metric $g_{\alpha\beta}$ this partition function does not have a universal non-ambiguous definition, due to the possibility to include non-universal local counterterms. This does not cause any troubles though, because as we will see, the path integral in (\ref{ZTT})  localizes on constant metrics.

Evaluation of the path integral (\ref{ZTT}) proceeds along the lines of classic textbook calculations of the  one-loop string amplitudes \cite{Polchinski:1998rq}.  There are some interesting differences, however.  
In particular,  we do not have Weyl gauge symmetry, so effectively we are dealing with non-critical strings. Another difference is that the integral is performed over vielbeins rather than over metrics. This brings in an additional integration variable---an overall angle $\phi$ of the vielbein. Indeed, reparametrization invariance allows to bring any metric $g_{\alpha\beta}$ into the canonical form 
\begin{equation}
\tilde{g}_{\alpha\beta} = e^{2\Omega(\sigma)} \left(
\begin{array}{cc}
1& \bar\tau_1  \\ 
\bar\tau_1& \bar\tau_1^2 +  \bar\tau_2^2
\end{array} 
\right)\;,
\label{gauge_fix_g}
\end{equation} 
where $\bar \tau=\bar \tau_1+i\bar\tau_2$ is the standard modular parameter and 
\[
0\leq\sigma^\alpha<|\Lambda|^{-1/2}\;.
\]
 are periodic  worldsheet coordinates.
This translates into the following expression for the dyad components
\begin{equation}
\tilde{e}^{a}_\alpha = e^{\Omega(\sigma)}\left(e^{\epsilon\phi(\sigma)}\right)^a_{\;\;b}\hat{e}^b_\alpha(\bar\tau)\;,
\label{gauge_fix_e}
\end{equation}
where the $\epsilon$-symbol in the exponent stands for the generator of internal $SO(2)$ rotations, so that $\phi(\sigma)$ parametrizes an arbitrary $\sigma$-dependent rotation in the internal space. Equation \eqref{gauge_fix_e} is the analogue on the torus of \eqref{eq_epmlc}. Recall that the internal  $SO(2)$ group is not a gauge symmetry in this description, so that $\phi(\sigma)$ is a genuine integration variable in the expression (\ref{ZTT}) for the partition function.
Here $\hat{e}^b_\alpha(\bar\tau)$ is a constant dyad of the form
\begin{equation}
\hat{e}^a_\alpha = \left(
\begin{array}{cc}
1& \bar\tau_1  \\ 
0& \bar\tau_2 
\end{array} 
\right)\;.
\end{equation}
To evaluate the partition function we will express the integration measure ${\cal D}e$ as an integral over $\Omega,\phi,\bar\tau$ and the worldsheet diffs $v$. Following the Faddeev--Popov (FP) method we insert a unity in the form 
\begin{equation}
1 = J(e)\int {\cal D}v{\cal D}\Omega{\cal D}\phi d^2\bar\tau \delta\left(e^{(v)} - \tilde{e}\right),
\label{FP_one}
\end{equation} 
where $e^{(v)}$ is the image of the dyad $e$ under the action of  $v$ and 
$J(e)$ is a diff invariant FP determinant. 

Then the standard FP manipulations allow to factor out the ${\cal D} v$ integral, which cancels out with $V_{diff}$ in (\ref{ZTT}). The ${\cal D}e$ integral is now trivial due to the presence of a $\delta$-function in (\ref{FP_one}) and the dressed partition function turns into
\be
\label{ZTTFP}
Z_{JT} = \int {\cal D}\Omega {\cal D}\phi d^2\bar\tau {\cal D}X J(\tilde{e})e^{-{\Lambda\over 2}\int\epsilon^{\alpha\beta}\epsilon_{ab}\l\d_\alpha X^a -\tilde e^a_\alpha\r\l\d_\beta X^b- \tilde e^b_\beta \r}
Z_0(\tilde{g}_{\alpha\beta})\;.
\ee
The next step is to perform the ${\cal D}X$ integral. The presence of windings implies that the target space embeddings $X^a$ can be decomposed as 
\begin{gather}
\label{wind}
X^a=|\Lambda|^{1/2}L^a_\alpha\sigma^\alpha+Y^a\;,
\end{gather}
where
\[
L^a_\alpha=(L_\alpha,L'_\alpha)
\]
and  $Y^a(\sigma)$ are scalar fields periodic on the worldsheet. They take values on a target space torus shown in Fig~\ref{fig:ABCD}. It is straightforward now to evaluate the integral 
\[
I =\int {\cal D}Xe^{\Lambda\int\epsilon^{\alpha\beta}\epsilon_{ab}\d_\alpha X^a \tilde{e}^b_\beta}\;,
\]
 which is  a part of (\ref{ZTTFP}).
It reduces simply to 
\be
\label{DX}
I =e^{\Lambda |\Lambda|^{1/2}\int \epsilon^{\alpha\beta}\epsilon_{ab}
L^a_\alpha  \tilde{e}^b_\beta}\int {\cal D}Ye^{-\Lambda\int\epsilon^{\alpha\beta}\epsilon_{ab} Y^a \d_\alpha\tilde{e}^b_\beta}=
{|\Lambda|{\cal A}\tilde{\cal A}\over (2\pi)^2}
e^{\Lambda |\Lambda|^{1/2}\int \epsilon^{\alpha\beta}\epsilon_{ab}
L^a_\alpha  \tilde{e}^b_\beta}
\delta\left(\frac{\epsilon^{\alpha\beta}}{\sqrt{\tilde g}}\partial_\alpha \tilde{e}^a_{\beta}\right)\;,
\ee
where ${\cal A}$ is the area of the target space torus (\ref{calA}), and $\tilde{\cal A}$ is the area of the worldsheet torus,
\[
\tilde{\cal A}=\int \sqrt{\tilde{g}}\;.
\]
The  factor of
\be
\label{AA}
{|\Lambda|{\cal A}\tilde{\cal A}\over (2\pi)^2}\;
\ee
 in (\ref{DX}) is the constant mode contribution (see Appendix~\ref{app:gory1} for the details on the integration measure).  
 The $\delta$-function constraint in (\ref{DX})
 \[
  \epsilon^{\alpha\beta}\d_\alpha\tilde{e}^b_\beta=0
 \]
 can only be satisfied for constant vielbeins $\tilde{e}^b_\beta$. As a result, after decomposing $\Omega$ and $\phi$ as a sum
 of  constant modes and orthogonal pieces
 \begin{gather}
 \Omega(\sigma)={\bar\Omega}+\Omega'(\sigma)\\
 \phi(\sigma)={\bar\phi}+\phi'(\sigma)
 \end{gather}
 the $\delta$-function in (\ref{DX}) takes the following form
\begin{equation}
\label{deltapr}
\delta\left(\frac{\epsilon^{\alpha\beta}}{\sqrt{g}}\partial_\alpha \tilde{e}^a_{\beta}\right) = 
\frac{\delta(\Omega')\delta(\phi')}{\sqrt{\det'{Q^\dagger(\tilde e) Q(\tilde{e})}}}
\end{equation}
where $Q$ is the differential opperator defined by its action on infinitesimal variations of $\Omega'$ and $\phi'$ as
\begin{equation}
\label{Qdef}
Q(e)\left(\begin{array}{c}
\delta\Omega'\\ 
\delta\phi'
\end{array} \right) = \frac{1}{\sqrt{g}}e^a_\beta \epsilon^{\beta\alpha}\partial_\alpha\delta\Omega' + \frac{1}{\sqrt{g}}\epsilon^a_{\;\;b}e^b_\beta \epsilon^{\beta\alpha}\partial_\alpha\delta\phi'\;.
\end{equation}
  and $\det'$ stands for the product of non-vanishing eigenvalues.
  As a result the dressed partition function reduces to
  \be
  \label{almost}
  Z_{JT}={{\cal A}e^{-\Lambda{\cal A}}\over (2\pi)^2}\int_{-\infty}^\infty d\bar{\Omega}\int_0^{2\pi}d\bar{\phi}\int_Pd^2\bar\tau {\l\Lambda\bar{\cal A}\r^2
  J(\bar{e})\over
\sqrt{ \det'{Q^\dagger(\bar e) Q(\bar e)}} }
  e^{{\Lambda\over|\Lambda|}\epsilon^{\alpha\beta}\epsilon_{ab}\l\sqrt{|\Lambda|}
L^a_\alpha  \bar{e}^b_\beta  -{1\over 2}\bar{e}^a_\alpha \bar{e}^b_\beta \r}Z_0(\bar{g}_{\alpha\beta})\;,
  \ee
 where we replaced $\tilde{e}\to \bar{e}$, $\tilde{g}\to \bar{g}$ to indicate that the integral is now taken over the constant vielbeins only.
 The factor of $e^{-\Lambda{\cal A}}$ comes from the total derivative term (\ref{dXdX}).
  Note that we obtained an extra factor of $|\Lambda|\bar{\cal A}$ from the integration measure (see Appendix~\ref{app:gory1}). Also, due to the presence of windings in 
 (\ref{wind}) the integration over the modular parameters is not restricted to the $SL(2,\mathbb{Z})$ fundamental region, but extends to the whole upper half plane (or, equivalently, the Poincar\'e disc $P$), $-\infty<\bar\tau_1<\infty$ and $\bar\tau_2>0$, c.f.~\cite{Polchinski:1985zf,Dubovsky:2017cnj} . 
 
 The last remaining step is to evaluate the ratio of determinants in (\ref{almost}). This is performed along the lines of a similar calculation in the critical string case \cite{Polchinski:1985zf}. We defer the details of this straightforward but somewhat tedious calculation till Appendix~\ref{app:gory2}. 
 The result is quite simple
 \be
 \label{JQQ}
 {J(\bar{e})\over
 \sqrt{\det'{Q^\dagger(\bar e) Q(\bar e)}} }={1 \over \bar{{\cal A}}\bar\tau_2^2}\;,
 \ee
so that the final answer for the JT partition function is
 \be
  \label{Ztau}
  Z_{JT}={|\Lambda|{\cal A}e^{-\Lambda{\cal A}}\over (2\pi)^2}\int_{-\infty}^\infty d\bar{\Omega}e^{2\bar{\Omega}}\int_0^{2\pi}d\bar{\phi}\int_P{d^2\bar\tau\over\bar \tau_2} 
  e^{ {\Lambda\over|\Lambda|}\epsilon^{\alpha\beta}\epsilon_{ab}\l\sqrt{|\Lambda|}
L^a_\alpha  \bar{e}^b_\beta  -{1\over 2}\bar{e}^a_\alpha \bar{e}^b_\beta \r}Z_0(\bar{g}_{\alpha\beta})\;.
  \ee
 \subsection{Equivalence to the $T\bar{T}$ Deformed Spectrum and Localization}
 The answer (\ref{Ztau}) simplifies when rewritten as an integral over constant  dyads,
\[
\bar{e}^a_\alpha\equiv\sqrt{|\Lambda|}{\bar L}'^a_\alpha\equiv\sqrt{|\Lambda|}({\bar L}_\alpha,{\bar L}'_\alpha)
\]
and turns into
 \be
  \label{ZL}
  Z_{JT}={\Lambda^2{\cal A}e^{-\Lambda{\cal A}}}\int_{\bar{\cal A}>0}{d^4 \bar{L}\over (2\pi)^2 \bar{\cal A}}  e^{\Lambda \epsilon^{\alpha\beta}\epsilon_{ab}\l
L^a_\alpha  \bar{L}^b_\beta  -{1\over 2}\bar{L}^a_\alpha \bar{L}^b_\beta \r}Z_0(\bar{g}_{\alpha\beta})\;.
  \ee
When written in this form it is immediate to see the relation of the partition function $Z_{JT}$ to the partition function of the $T\bar{T}$ deformed theory.
Namely, 
 we recognize in $Z_{J T}$ a solution of the initial value problem for the linear diffusion equation (\ref{cardy}), written as a convolution with the heat kernel (c.f. Ref.~\cite{Cardy:2018sdv}). 
 The ``time" parameter of the diffusion equation is given by (\ref{ell}).

 However, the diffusion equation (\ref{cardy}) has many different solutions and most of them do not have the form expected from a partition function. So a bit of extra work is required to check that the solution (\ref{ZL}) is the correct one.
 In fact, the proper interpretation of the integral in (\ref{ZL}) is rather subtle. Indeed, the operator in the r.h.s. of the ``diffusion" equation (\ref{cardy}) is not elliptic. Related to that, the expression in the exponential in (\ref{ZL}) is not sign definite, and the integral (\ref{ZL}) is not well-defined as written for either signs of $\Lambda$.
 This should not come as a surprise. Path integral in Euclidean quantum gravity suffers from an infamous sign problem, and this is exactly what we find here. Let us, however, proceed under the assumption that the integral is properly regulated by an appropriate choice of the integration contours. 
 
 One possibility is to simply use equation (\ref{cardy}) with initial conditions given by $Z_0$ as a definition of integral (\ref{ZL}). To make this definition unambiguous it is necessary to specify boundary conditions at the surface of zero area ${\cal A}=0$. This is done most conveniently if the target-space torus is parametrized by the
 $(\phi,\Omega,\tau)$ variables, defined in the same way as in \eqref{gauge_fix_e}, see also Fig.~\ref{fig:ABCD}. Here $\phi$ and $\Omega$ are constant, not to be confused with the constant parts of $\phi(\sigma)$ and $\Omega(\sigma)$, parametrizing the worldsheet torus, which are denoted by $\bar\phi$ and $\bar\Omega$.  Then equation (\ref{cardy}) translates into the following equation for the partition function
\be
\label{eqoft}
\d_{\ell^2}Z_{JT}={e^{-2\Omega}\over 2}\l\d_\Omega\d_{\tau_2}-\d_\phi\d_{\tau_1}-\tau_2\l\d_{\tau_1}^2+\d_{\tau_2}^2\r-\tau_2^{-1}\d_\Omega  \r Z_{JT}\;,
\ee
and the boundary condition has to be imposed at $\tau_2=0$. To choose the right condition we inspect  the solution (\ref{Zsumapp}) discussed in Appendix~\ref{app:diffusion}, which in these variables turns into
 \be
 Z_{JT}=\sum_n e^{-{\tau_2 R}  E_n(R,\ell^2)+ 2\pi ik_n\tau_1}\;,
 \label{ZA}
 \ee
 where 
 \[
 R={|\ell|\over\sqrt{2}}e^\Omega\;.
 \]
 It is immediate to see that (\ref{ZA}) satisfies
\be
\d_\Omega Z_{JT}\Big |_{\tau_2=0}=0 \;,
\label{NBC}
\ee
 which is nothing but Neumann boundary condition, since at $\tau_2=0$ the second derivative part of (\ref{eqoft}) has the form $\d_\Omega \d_{\tau_2}$. We thus conclude that one way to define the JT partition function is through equation (\ref{cardy}) or (\ref{eqoft}) supplemented with boundary condition (\ref{NBC}). 
As a note of caution, let us stress that for a generic initial condition $Z_0$ the solution (\ref{ZL}) is  different from the one satisfying the Neumann's boundary condition. To see
that for initial conditions corresponding to a physical partition function (given by (\ref{ZA}) at $\ell^2=0$)  
the solution (\ref{ZL}) does satisfy (\ref{NBC}) and reproduces (\ref{ZA}) at all $\ell^2$ let us proceed with a more direct evaluation of  (\ref{Ztau}).
 
 Namely, we will see now that a saddle point evaluation of the integral, which turns out to be one-loop exact, leads directly to a solution of the hydrodynamical equation (\ref{burgers}), bypassing the diffusion equation for the partition function.  As we already stated, the integral is formally divergent, so that at each step we will be making the most optimistic (and, hopefully, reasonable) assumptions about the integration contours.
  It turns out this computation is most conveniently done using a hybrid set of variables between $(L, L')$ and $(\phi, \Omega, \tau)$. Namely, we will use
 $(L,\tau)$ as a set of variables parametrizing the target space torus (\ref{LLptorus}). Note that $(\Omega, \phi)$ are essentially polar coordinates in the $L$ plane,
 \begin{gather}
 L_1={|\ell|\over \sqrt{2}}e^\Omega\cos\phi\;,\\
 L_2={|\ell|\over \sqrt{2}}e^\Omega\sin\phi\;.
 \end{gather}
 Analogous set of coordinates, $(\bar L,\bar\tau)$, will be used as the integration variables.
 Then combining (\ref{ZL}) and (\ref{ell}) we get
 \be
 \label{ZTT1}
Z_{JT}=\sum_n
{4{\cal A}\over (2\pi)^2\ell^4}e^{2{\cal A}/\ell^{2}}\int^\infty_{-\infty}d^2\bar L \int_P{d^2\bar\tau\over\bar\tau_2}e^{{2\over\ell^2} \l \bar R^2\bar\tau_2- R(\bar L_1(\bar\tau_2+\tau_2)+\bar L_2(\bar\tau_1-\tau_1))\r}e^{-{\bar\tau_2 \bar R}  E_n(\bar R,0)+ 2\pi ik_n\bar\tau_1}\;,
 \ee 
 where $R$, $\bar R$ are defined as in (\ref{Rdef}) and we used the standard expression
 \be
\label{Z0}
Z_0=\sum_n e^{-T  E_n(R,0)+ i{ P_n} \Delta }
\ee
 for the undeformed partition function.
 Note, that in principle one can make a change of variables $\bar\tau_1\to\bar\tau_1+\tau_1$ which makes it explicit that each of the terms in  the sum for $Z_{JT}$ has the correct $\tau_1$ dependence expected for a partition function ({\it i.e.}, proportional to 
 $e^{2\pi ik_n\tau_1}$).
 Let us first consider the case when the undeformed theory is a CFT, so that
 \[
 E_n(\bar R,0)={{\cal E}_n\over \bar R}\;.
 \] 
 with some constant ${\cal E}_n$'s.
 In this case the $\bar L$ integral is Gaussian and can be readily evaluated with the following result,
 \be
 \label{ZCFT}
 Z^{CFT}_{JT}=\sum_n
 {{\cal A}\over 2\pi\ell^2}e^{2{\cal A}/\ell^{2}}\int_P{d^2\bar\tau\over\bar\tau_2^2}e^{-{R^2\over 2\ell^2\bar\tau_2}((\bar\tau_1-\tau_1)^2+(\tau_2+\bar\tau_2)^2)}e^{-{\bar\tau_2 {\cal E}_n+ 2\pi ik_n\bar\tau_1}}\;.
 \ee
 The same integral representation was obtained in \cite{Dubovsky:2017cnj} for a $T\bar{T}$ deformed critical $c=24$ CFT using the Polyakov formalism. We see that the JT description  extends this result to non-critical CFT's\footnote{To be precise, by the Polyakov formalism we mean here extending the theory by introducing a coupling to the metric and two minimally coupled massless bosons. It is plausible that the $T\bar{T}$ deformed spectrum for a general CFT can be obtained by replacing one of the minimally coupled bosons with a linear dilaton, similar to, e.g., \cite{Dodelson:2017emn}. At the moment the relation between the JT and linear dilaton formalisms is unclear.}.
 Integration over the modular parameters is straightforward and leads to the expected result 
 \be
 Z^{CFT}_{JT}=\sum_ne^{2\pi ik_n\tau_1-\tau_2R\ell^{-2}
\l \sqrt{R^2+2\ell^2{\cal E}_n+{4\pi^2 k^2\ell^4\over R^2}}-R\r}\;.
 \ee
 A notable property of the modular integral in (\ref{ZCFT}) is that it is localizable -- $Z^{CFT}_{JT}$ is dominated by the contribution of a single saddle point and is one-loop exact. As explained in \cite{Dubovsky:2017cnj} this property can be understood by applying the Duistermaat--Heckman localization formula \cite{Duistermaat:1982vw} to the Poincar\'e disc. Given that the $\bar L$ integration in (\ref{ZTT1}) is Gaussian we find that the localization property of the 
 $T\bar{T}$ deformed partition function extends to  a general CFT case. 
 
 Coming to the general QFT case note first that at early ``times" at $\ell^2\to 0$ the integral in (\ref{ZTT1}) is dominated  by a saddle, which is insensitive to the undeformed energies. At this saddle the worldsheet torus coincides with the target space torus
 \[
 \bar L^a_\alpha=L^a_\alpha\;,
 \]
 which enforces the correct initial condition
 \[
 Z_{JT}(\ell^2\to 0)\to Z_0\;.
 \]
 
Let us now take a closer look at the integral  in (\ref{ZTT1})
and check that $Z_{JT}$ is one-loop exact also for a general quantum field theory.
First, note that the $\bar\tau_1$ integration results in a $\delta$-function fixing $\bar L_2$ to a
 constant value. Hence, integration over the $(\bar\tau_1,\bar L_2)$ pair is one-loop exact and gives
 \be
 \label{hbar}
Z_{JT}=\sum_ne^{2\pi ik_n\tau_1}{{\cal A}\over \pi\ell^2R}e^{2{\cal A}/\ell^{2}}
\int^\infty_{-\infty}d\bar L_1 \int_0^\infty{d\bar\tau_2\over\bar\tau_2}
e^{{2\over\hbar\ell^2} \l \bar R^2\bar\tau_2-
 R\bar L_1(\bar\tau_2+\tau_2)\r}e^{-{1\over\hbar}{\bar\tau_2 \bar R}  E_n(\bar R,0)}\;,
 \ee 
where
\[
\bar R=\sqrt{\bar L_1^2-{\pi^2\ell^4 k_n^2\over R^2}}\;.
\]
We introduced a fictitious Planck constant $\hbar$ in (\ref{hbar}) to emphasize that we are evaluating this integral via saddle point. At the end we set
$\hbar=1$. 
Evaluating (\ref{ZTT1}) in the one-loop approximation we get
\be
\label{1loop}
Z_{JT}=\sum_ne^{2\pi ik_n\tau_1}e^{{2\over\ell^{2}}\l{\cal A}-R\bar L_{1}\tau_2\r}\;,
\ee
where $ \bar L_{1}$ is now a solution to the saddle point equation,
\be
\label{L1s}
F(\bar L_1)\equiv{2\over\ell^2}\l\bar R^2-R\bar L_{1}\r-\bar R E_n(\bar R,0)=0\;.
\ee
From (\ref{1loop}) we deduce that
\be
\label{EnlR}
E_n(R,\ell^2)={2\over \ell^2}\l \bar L_1-R\r\;.
\ee
It is straightforward to check that (\ref{EnlR}) is indeed a solution of (\ref{burgers}) provided the saddle point equation (\ref{L1s}) is satisfied.
In particular, at $k=0$ eqs.~(\ref{EnlR}) and (\ref{L1s})  combine into the well-known implicit solution to the inviscid Burgers' equation \cite{Smirnov:2016lqw,Cavaglia:2016oda},
\[
E_n(R,\ell^2)=E_n\l R+{\ell^2\over 2}E_n(R,\ell^2),0\r\;,
\]
where we picked the branch of solutions, which has a smooth  $\ell^2\to 0$ limit in (\ref{L1s}). We see that the saddle point approximation
to (\ref{hbar}) indeed reproduces the expected exact answer. 

This strongly suggests that the integral (\ref{hbar}) is one-loop exact for a general energy spectrum. As a further support for this expectation we checked
that the leading order correction in the semiclassical expansion of (\ref{hbar}) indeed vanishes. In principle, this calculation is straightforward to push to higher orders, but it rapidly becomes rather tedious and not very illuminating. Instead, let us   
 perform a few more formal operations with (\ref{hbar}) which allow to evaluate it exactly and shed some light on where the
 one-loop exactness comes from. First, let us integrate by parts in $\bar L_1$ to get
\be
Z_{JT}=\sum_ne^{2\pi ik_n\tau_1}{e^{2{\cal A}/\ell^{2}}\over 2\pi}
\int^\infty_{-\infty}d\bar L_1 F'(\bar L_1) e^{-{2  \tau_2 R \bar L_1\over\ell^2}} \int_0^\infty{d\bar\tau_2}
e^{\bar \tau_2F(\bar L_1)}\;.
\ee
Now we can take the integral over $\bar\tau_2$, again being optimistic about the integration contour,
\be
\label{res}
Z_{JT}=\sum_ne^{2\pi ik_n\tau_1}{e^{2{\cal A}/\ell^{2}}\over 2 \pi}
\int^\infty_{-\infty}d\bar L_1 \frac{ i F'(\bar L_1)}{ F(\bar L_1)+i\epsilon} e^{-{2  \tau_2 R \bar L_1\over\ell^2}}.
\ee
The remaining integral over $\bar L_1$ can be taken by residues, localizing the answer on the solutions of (\ref{L1s}). When an 
unperturbed theory  is close to CFT there are two solutions of $F(\bar L_1)+i\epsilon=0$, one in the upper and one in the lower half-planes.
By closing the contour (\ref{res}) one picks up one of these contributions\footnote{If the theory has mass scales of order $\ell^{-1}$ there can be other poles corresponding to states with energies that are non-perturbative in $\ell$ and that are not present in the undeformed theory. In order to reproduce the $T\bar{T}$ spectrum one needs to chose the contour that avoids those poles.}. We see that the localization property of the deformed partition function is related to the possibility to calculate it using the residue theorem.

So we find that the $T\bar T$ deformed partition function is one-loop exact for any quantum field theory. 
This is a very satisfactory result given  the semiclassical nature of gravitational dressing (\ref{USU}), as discussed in detail in
\cite{Dubovsky:2017cnj}. In addition, it provides some justification for our treatment of the integral in (\ref{ZL}). Indeed, as we said, 
many of our manipulations are formal at best because as written for any sign of $\Lambda$ some of the saddle integrations  are performed over contours where the saddle is a minimum, rather than a maximum.  However, the localization property of these integrals at least provides a precise algebraic meaning to these manipulations. 

To conclude, let us come back to the question whether the first order description (\ref{TTaction}) is just technically more convenient compared to the conventional JT gravity action (\ref{action}), or it is really different in a finite volume. The consequence of using the vielbein formalism is the presence of the polar angle $\phi$ in the integral (\ref{ZTT1}) for the partition function. Of course, given that the non-gravitational part of (\ref{ZTT1})
is $\phi$ independent, it is straightforward to integrate over $\phi$. The result is 
\be
\nonumber 
Z_{JT}=\sum_ne^{2\pi ik_n\tau_1+2{\cal A}/\ell^{2}}{2{\cal A}\over \pi \ell^4}\int^\infty_0\bar Rd\bar R \int_P{d^2\bar\tau\over\bar\tau_2}
I_0\l x\r e^{{2\over\ell^2} \bar R^2\bar\tau_2-{\bar\tau_2 \bar R}  E_n(0,\bar R)+ 2\pi ik_n\bar\tau_1}\;,
 \ee 
where 
\[
x={2R\bar R\over \ell^2}\sqrt{\bar\tau_1^2+(\tau_2+\bar\tau_2)^2}
\]
and $I_0(x)$ stands for the modified Bessel function of the first kind. This expression depends only on the variables present in the original JT action (\ref{action}), but it is hard to see how to arrive at this integration measure without going through (\ref{TTaction}). We take this as an indication that the vielbein formalism is more than just a technical convenience in the present setup. This poses a question whether one should add to (\ref{Ztau}) also a contribution from the lower half plane in $\tau$ space,  to account for  the $Z_2$ factor in the $O(2)$ internal symmetry group. Given that our manipulations with this integral are quite formal at the point it is hard to decide on this. Perhaps this can be resolved by considering other geometries, or theories with fermions which are directly sensitive to the spin structure.

\section{Conclusions and Future Directions}
\label{sec:last}
To summarize, we presented a derivation of the finite volume spectrum corresponding to the gravitationally dressed $S$-matrix (\ref{USU}) by a brute force evaluation of the torus partition function in the flat space JT gravity. Admittedly, the paper came out quite technical and the presented derivation is longer,
less elegant and less inventive compared to the earlier ones \cite{Smirnov:2016lqw,Cavaglia:2016oda,Cardy:2018sdv} based on the $T\bar{T}$  line of reasoning. However, to large extent our main goal here was to show that after the proper action principle (\ref{TTaction}) is identified one does not need to be smart and inventive. All that is needed is to follow the standard rules of quantum field theory and to be careful in doing calculations\footnote{Of course, this is just a corollary of the well-known theorem that if you do things right you get the correct result. One of the authors is thankful to Valery Rubakov for teaching him this very important piece of knowledge among many others.}. 

This is somewhat non-trivial in the present context, given that we are dealing with a gravitational path integral rather than with a conventional quantum field theory. In fact, we do not feel that we fully succeeded, given a rather formal character of our manipulations with the integral (\ref{ZL}). 
The possibility to independently derive the same finite volume spectrum using the $T\bar T$ deformation arguments and, for integrable theories, using the TBA technique definitely adds to our confidence in the correctness of the JT description. Also, the localization property of the JT gravity provides
at least an exact algebraic meaning to the resulting Euclidean path integral. It will be interesting to study what happens for deformations of the whole setup which are not one-loop exact.

In the present paper we focused on calculating the torus partition function in the flat space JT gravity. The next natural step
 is to extend this analysis to other geometries. Several interesting results in this direction have already been obtained in \cite{Cardy:2018sdv}. As discussed in \cite{Dubovsky:2017cnj} an especially interesting and straightforward case to consider is the torus partition function with a winding  for only one of the ``target space" coordinates. This partition function does not have a direct interpretation within a two-dimensional theory. However, there are suggestive indications \cite{Dubovsky:2015zey,Dubovsky:2016cog,Dubovsky:2018dlk} that worldsheet theories of confining strings are closely related to the gravitational dressing. In particular, the worldsheet theory of three-dimensional gluodynamics was suggested to be in the same equivalence class as a single dressed massless boson, and the worldsheet theory of the massive adjoint QCD$_2$ may be a deformation of a dressed massless fermion. If correct, this implies that the dynamical coordinates $X^a$ indeed have the meaning of target space coordinates, and the above calculation may allow to extract the spectrum of short stings (glueballs) similarly to the fundamental string case \cite{Polchinski:1985zf}.
 
Conventionally,  following \cite{Polyakov:1981rd}, one associates non-critical strings with the presence of a dynamical Liouville mode. Currently, there is a considerable experimental (or, better to say, lattice) and theoretical evidence that confining strings in $D=4,3,2$ Yang--Mills theory follow a different path. JT gravity demonstrates that it is indeed possible to couple two-dimensional gravity to a non-critical CFT (and, more generally, to a non-conformal quantum field theory) without introducing local Liouville dynamics.  

In addition, it will be interesting to generalize the JT description to other relatives of the $T\bar{T}$ deformation. These include its higher spin versions 
\cite{Smirnov:2016lqw} as well as a non-relativistic $J\bar T$ deformation \cite{Guica:2017lia,Bzowski:2018pcy}. It is natural to expect that 
similarly to $T\bar T$ these may be described by introducing a topological gauging of the corresponding symmetries.
Also by now there are examples of two-dimensional theories distinct from the $T\bar{T}$ deformation but with a similar UV behavior. These arise in  little strings \cite{Giveon:2017nie} and also as worldsheet theories in $QCD_2$ \cite{Dubovsky:2018dlk}. Hopefully, the JT description can be extended 
and will prove to be useful to describe these setups as well.

\section*{Acknowledgements} 
We thank  Mehrdad Mirbabayi for collaboration on related topics and for numerous fruitful discussions.
We are grateful to John Cardy, Chang Chen, Volodya Kazakov, Uri Kol, Massimo Porrati, Eva Silverstein and Shimon Yankielowicz  for many helpful discussions and correspondence. 
This work is supported in part by the NSF CAREER award PHY-1352119.

\appendix
\section{Direct Derivation of the Diffusion Equation}
\label{app:diffusion}
Let us present here a direct derivation of the diffusion equation (\ref{cardy}) for the partition function from a non-linear ``hydrodynamical" equation (\ref{burgers}) for the energy levels. This derivation is essentially the same as the one presented in the Appendix of an earlier version of \cite{Cardy:2018sdv}, apart from a slightly different choice of variables. The final result contains an additional factor of area ${\cal A}$ in (\ref{psi}) as compared to an earlier version of \cite{Cardy:2018sdv}.

Consider a deformed theory on a torus shown in Fig.~\ref{fig:ABCD}.
We treat 
\be
\label{Rdef}
AB\equiv R =\sqrt{L_1^2+L_2^2}
\ee
 as the spatial direction. Then  the partition function reads as
\be
\label{Zsumapp}
Z=\sum_n e^{-T  E_n(R,\ell_s^2)+ i{ P_n} \Delta }
\ee
where the sum goes over all energy levels $E_n(R,\ell_s^2)$ and 
\[
P_n={2\pi k_n\over R}
\]
 are the corresponding momenta, so that $k_n$ are integers.
Here
\[
T={{\cal A}\over R}={L_1L'_2-L'_1L_2\over R}
\]
is the Euclidean  time periodicity and
\[
\Delta={\sqrt{ R^2(L_1^{'2}+L_2^{'2})-{\cal A}^2}\over R}={L_1L'_1+L_2L'_2\over R}
\]
determines the twist of a torus, see Fig.~\ref{fig:ABCD}.
Plugging these expressions into  (\ref{Zsumapp}) we obtain 
\[
Z=\sum_ne^{-{{\cal A}\over R}{E}(R,\ell_s^2)+2\pi i k{L_1L'_1+L_2L'_2\over R^2}}\;.
\]
By acting on this expression with the (corrected) Cardy operator
\[
{{ \cal D}}^2={1\over 2}\l\d_{L_1}\d_{L'_2}-\d_{L_2}\d_{L'_1}-{\cal A}^{-1} (L_1\d_{L_1}+L_2\d_{L_2}+L'_1\d_{L'_1}+L'_2\d_{L'_2})\r
\]
we obtain
\be
{{\cal D}}^2Z={{\cal A} Z\over 2R^2}
\sum_n \l{ P_n}^2+R{ E_n}\d_R{ E_n}\r\;.
\ee
Also
\be
\d_{\ell_s^2}Z=-{{\cal A} Z\over R}\sum_n\d_{\ell_s^2}{ E_n} \;.
\ee
Hence we find that the hydrodynamical equation (\ref{burgers})
implies a linear equation for the partition function
\be
\d_{\ell_s^2}Z=-{{\cal D}}^2Z\;.
\ee
It is immediate to check that it takes the form (\ref{cardy}) after the substitution (\ref{psi}). Note that this equation holds separately for each individual term in the sum (\ref{Zsumapp}). 

\section{Scalar Field Integration on a Torus}
\label{app:gory1}
In this Appendix and the following one we provide details on the integration measure in the JT path integral and on the derivation of the Jacobians $J$ and $\det'Q^\dagger Q$ in  (\ref{almost}). To define the integration measure we need to introduce a metric (c.f., 
\cite{Polyakov:1981rd,Polchinski:1985zf,DHoker:1988pdl}) in the field space. By metric we mean here an inner product in the tangent space.
For scalar fields (such as $Y^a$, $\Omega$ and $\phi$) the metric $G_s$ is given by 
\be
\label{scmetric}
G_s(\delta \psi, \delta \chi)=|\Lambda|\int \sqrt{g}\delta \psi\delta \chi\;,
\ee
where $\delta \psi$, $\delta \chi$ are  infinitesimal field deformations.
This expression explains the origin of the normalization factor (\ref{AA}) and of the additional $\bar{A}$ factor in (\ref{almost}).
Namely, to  derive (\ref{DX}) one writes
\be
\int {\cal D}Y e^{-\Lambda\int\epsilon^{\alpha\beta}\epsilon_{ab} Y^a \d_\alpha\tilde{e}^b_\beta}=I_0I_1\;,
\ee
where we separated $Y^a$ into a sum of a constant mode and an orthogonal complement,
\[
Y^a=\bar{Y}^a+Y'^a\;
\]
so that $I_0$ is the constant mode contribution and $I_1$ is all the rest. We have then
\be
\label{I0}
I_0=\int d^2 \bar{Y}\l\sqrt{\det\bar{G_s}}\r^2=|\Lambda|{\cal A}\bar{\cal A}\;,
\ee
where $\bar{G}_s$ is a restriction of the metric (\ref{scmetric}) on a constant mode subspace. For $I_1$ we get {\bf $\tilde g$??}
\be
\label{I1}
I_1=\int {\cal D}Y' e^{-\Lambda\int\epsilon^{\alpha\beta}\epsilon_{ab} Y'^a \d_\alpha\tilde{e}^b_\beta}=
\int {\cal D}Y' e^{ -G_s(Y'^a,{\epsilon^{\alpha\beta}\over \sqrt{g}}\epsilon_{ab}  \d_\alpha\tilde{e}^b_\beta)}=\l\mbox{\rm det}'{\,2\pi}\r^2 \delta\left(\frac{\epsilon^{\alpha\beta}}{\sqrt{\tilde g}}\partial_\alpha \tilde{e}^a_{\beta}\right)
\ee
where $\det'$ stands for the product over all non-constant modes. This factor comes from the standard expression for the $\delta$-function
\[
\int dy e^{-i y p}={2\pi}\delta(p)\;.
\]
To evaluate $\mbox{\rm det}'{\,2\pi}$ note that
\be
\label{detp}
\mbox{\rm det}'{\,2\pi}={1\over 2\pi}\mbox{\rm det}{\,2\pi}={1\over 2\pi}\;
\ee
where at the last step we made use of the fact that a determinant of a multiplication by a constant $\det C$ can be absorbed into a renormalization of the vacuum energy. Combining (\ref{I0}), (\ref{I1}) and (\ref{detp}) we reproduce the normalization factor (\ref{AA}).

Let us point out two subtleties in this calculation. First, strictly speaking the integral in (\ref{I1}) gives $\delta$-function only for imaginary values of $\Lambda$ and is not well-defined as written. In principle, this is a common situation encountered 
when one performs a Wick rotation of integrals with Lagrange multipliers $\lambda$. The natural way out seems to analytically continue also the integration contour $\lambda\to i\lambda$. The apparent subtlety is that in the present case we want the constant modes $\bar Y^a$ to span a Euclidean torus. Hence, this Wick rotation needs to be performed only for $Y^{'a}$ rather than for the full $Y^a$.

Second,  the $\delta$-function in (\ref{I1}) is not a complete one---it acts only on the non-constant subspace $Y'$. However,
its argument in (\ref{I1}) is a total derivative, so it does not contain constant modes, which is consistent with 
 the way we use it in (\ref{deltapr}).

Similarly, an additional factor of $|\Lambda|\bar{\cal A}$ in (\ref{almost}) comes from separating the contribution of constant modes $\bar{\Omega}$, $\bar{\phi}$ in the measure,
\[
{\cal D}\Omega{\cal D}\phi={d\bar\Omega }{d}\bar\phi \l\sqrt{\det \bar G_s}\r^2{\cal D}\Omega'{\cal D}\phi'\;.
\]
\section{Determinants}
\label{app:gory2}
The remaining step is to evaluate the determinant ratio (\ref{JQQ}). Fortunately, our task is simplified because we only need this ratio for constant metrics. Let us start with the simpler determinant $\sqrt{\det'{Q^\dagger(\bar e) Q(\bar e)}} $.
Using the definition (\ref{Qdef}) of $Q$ one finds
\begin{equation}
Q^\dagger Q = \left(\begin{array}{cc}
-\bar g^{\alpha\beta}\partial_\alpha\partial_\beta&0  \\ 
0&-\bar g^{\alpha\beta}\partial_\alpha\partial_\beta 
\end{array} \right)
\end{equation}
so that 
\begin{equation}
\label{QQ}
\sqrt{
\mbox{\rm det}' Q^\dagger Q}
= \mbox{\rm det}'\left(-\d^2\right). 
\end{equation}
Evaluation of the FP factor $J(\bar e)$ is more involved. In addition to the metric on a space of scalar fields (\ref{scmetric}) we will need now a metric on vielbeins $G_e$, diffs $G_v$ and moduli $G_\tau$. These are given by 
\begin{gather}
G_{e}(\delta e,\delta f)   =|\Lambda| \int \sqrt{g} g^{\alpha\beta}\delta_{ab}\delta e^a_\alpha\delta f^b_\beta\;,\\
G_{v}(\delta v,\delta u)   = |\Lambda|\int \sqrt{g} g_{\alpha\beta}\delta v^\alpha\delta u^\beta\;,\\
\label{gtau}
G_{\bar\tau}(\delta \bar\tau,\delta \bar\rho)=\delta^{ij}\delta\bar\tau_i\delta\bar\rho_j\;.
\end{gather}
Here $G_e$ and $G_v$ are determined by reparametrization invariance, and $G_{\bar\tau}$ is determined by $G_e$ from the 
embedding\footnote{A specific definition of $G_{\bar\tau}$ is not important as soon as it is consistently used everywhere.} 
$(\Omega,\phi,\bar\tau)\hookrightarrow e^a_\alpha$. 
To calculate $J(\bar e)$ note first that the $\delta$-function in (\ref{FP_one}) has a non-trivial support on a two-dimensional surface ${\cal M}$ rather than at an isolated point. This is the familiar statement that the conformal gauge condition does not fully fix gauge freedom on a torus. For a constant dyad $e=\bar e$ this surface is simply 
\[
{\cal M}(\bar e)=(v_T,\bar\Omega,\bar\phi,\bar\tau)\;,
\]
where $v_T$ is an arbitrary constant translation. Decomposing the ${\cal D}v$ integration measure as
\[
{\cal D}v={\cal D}v_T{\cal D}v'\;,
\]
where $v'$ is an orthogonal complement to $v_T$, we obtain,
\be
J^{-1}(\bar e)=\int{\cal D}v_T\int {\cal D}v'{\cal D}\Omega {\cal D}\phi d^2\bar\tau \delta\left(\bar e^{(v)} - \tilde{e}\right)={\int{\cal D}v_T\over \sqrt{\det'\l P^\dagger(\bar e)P(\bar e)\r}}\;,
\ee
where 
\begin{equation}
P(\bar e)\left(
\begin{array}{c}
\delta v\\ 
\delta\Omega\\ 
\delta\phi\\ 
\delta \bar\tau
\end{array} 
\right) = 
\bar{e}^{a}_\beta\partial_\alpha \delta v^\beta -  \delta\Omega \bar{e}^a_\alpha - \epsilon^a_{\;\,b}\bar{e}^b_\alpha\delta\phi -
\partial_{\bar\tau_i}\bar{e}^a_\alpha\delta\bar\tau_i\;.
\end{equation} 
As before $\det'$ indicates that the determinant is calculated over an orthogonal complement to the tangent space of ${\cal M}$.
For the area of ${\cal M}$ we get 
\be
\label{Marea}
\int{\cal D}v_T=\int_0^{|\Lambda|^{-1/2}}dv_T^1\int_0^{|\Lambda|^{-1/2}}dv_T^2\sqrt{\det {G_{v_T}}}=|\Lambda|{\bar{\cal A}^2}\;,
\ee
where $G_{v_T}$ is the restriction of the metric $G_v$ on the subspace of constant translations.
 
 The only remaining but somewhat cumbersome step is to calculate $\mbox{\rm det}'(P^\dagger P)$, where
 \begin{equation}
P^\dagger P
\l\begin{array}{c}
\delta v^\beta\\
\delta\Omega\\
\delta\phi\\
\delta\bar\tau_j
\end{array}
\r
 = \left(
\begin{array}{cccc}
-\delta^\alpha_\beta \partial^2 & \partial^\alpha  & {s}^\alpha_\gamma\partial^\gamma & 0 \\ 
-\partial_\beta & 2  & 0   &{k}_i  \\ 
s^\gamma_\beta\partial_\gamma&0  &2  &{f}_i  \\ 
0&k_i\mathcal{P}  & f_i \mathcal{P}  & A_{ij}
\end{array} 
\right)
\l\begin{array}{c}
\delta v^\beta\\
\delta\Omega\\
\delta\phi\\
\delta\bar\tau_j
\end{array}
\r
\end{equation}
with
\begin{gather}
s^\alpha_\gamma  = \sqrt{\bar g}\epsilon_{\beta\gamma}\bar g^{\beta\alpha}\\
k_i = \bar e^\alpha_a\d_{\bar\tau_i}\bar e^a_\alpha
\\
f_i=\epsilon_{ab}\d_{\bar\tau_i}\bar e^a_\alpha \bar e^{b\alpha} 
\\
A_{ij} =| \Lambda|\bar{\cal A}\bar g^{\alpha\beta}\partial_{\bar\tau_i}\bar e_{a\alpha}\partial_{\bar\tau_j}\bar e^a_\beta
\end{gather}
and ${\cal P}$ is an averaging operator
\[
{\cal P}\delta \phi=|\Lambda|\int \sqrt{\bar g}\delta\phi\;.
\]
Following the same strategy as in \cite{Polchinski:1985zf} we present  $P^\dagger P$ as the following product
\begin{equation}
P^\dagger P = T^\dagger R^\dagger N R T 
\end{equation} 
where $N$ is diagonal
\begin{equation}
N = \left(
\begin{array}{cccc}
-{1\over 2}\delta^\alpha_\beta\partial^2 
& 0 &0  &0  \\ 
0& 2 & 0 &0  \\ 
0&0  &2  &0  \\ 
0&0  &0  & {|\Lambda|\bar{\cal A}\over 2\bar\tau_2^2}\delta_{ij} 
\end{array} 
\right)\;,
\end{equation}
and the other matrices are triangular
\begin{gather}
T =\left(
\begin{array}{cccc}
1&0  &0  &0  \\ 
-\frac{1}{2}\partial_{\beta}&1  &0  &0  \\ 
\frac{1}{2}s^\gamma_\beta\partial_\gamma&  0&1  &0  \\ 
0&0  &0  &1 
\end{array} 
\right)\;,\\
R =\left(
\begin{array}{cccc}
1&0  &0  &0  \\ 
0&1  &0  &\frac{1}{2}{k}_i  \\ 
0&0  &1  &\frac{1}{2} {f}_i  \\ 
0& 0 & 0 & 1
\end{array} 
\right)\;.
\end{gather}
As a result, we obtain
\be
\label{detPP}
\sqrt{\mbox{\rm det}' P^\dagger P}=\sqrt{\mbox{\rm det}' N}={|\Lambda|\bar{\cal A}\over 2\bar\tau_2^2} \mbox{\rm det}' \l -{1\over 2}\partial^2\r  \det{2}=
{|\Lambda|\bar{\cal A}\over \bar\tau_2^2} \mbox{\rm det}' \l -\partial^2\r\;,
\ee
where we again used that $\det{2}$ can be absorbed in a renormalization of the vacuum energy. Combining (\ref{detPP}) with (\ref{Marea}) we obtain
\be
J={1\over  \bar{\cal A}\bar\tau_2^2} \mbox{\rm det}' \l -\partial^2\r\;,
\ee
which together with (\ref{QQ}) implies (\ref{JQQ}).
\bibliographystyle{utphys}
\bibliography{dlrrefs}
\end{document}